\documentclass[sigconf, 10pt, letter]{acmart}

\renewcommand\footnotetextcopyrightpermission[1]{} 
\usepackage{comment}
\usepackage{textcomp}
\usepackage{courier}
\usepackage{amsmath,amssymb,amsfonts}
\usepackage{algorithmic}
\usepackage{graphicx}
\usepackage{xcolor}
\usepackage{enumitem}
\usepackage{subfigure}
\usepackage{url}
\usepackage{hyperref}
\usepackage{comment}
\usepackage{textcomp}
\usepackage[T1]{fontenc}
\usepackage{textcomp}
\usepackage{filecontents}

\newcommand{\deepmac}{\texttt{DeepMAC}}

\def\BibTeX{{\rm B\kern-.05em{\sc i\kern-.025em b}\kern-.08em
    T\kern-.1667em\lower.7ex\hbox{E}\kern-.125emX}}

\usepackage{booktabs} 
\usepackage{paralist}

\usepackage[ruled]{algorithm2e}

\usepackage[colorinlistoftodos]{todonotes}

\begin{document}

\title{Unboxing MAC Protocol Design Optimization Using Deep Learning} 
\author{Hannaneh Barahouei Pasandi, Tamer Nadeem}

\affiliation{%
  \institution{Dept. of Computer Science, Virginia Commonwealth University, Richmond, VA 23284, USA}
}
\email{barahoueipash, tnadeem@vcu.edu}

\begin{abstract}
Evolving amendments of  802.11 standards feature a large set of physical and MAC layer control parameters to support the increasing communication objectives spanning application requirements and network dynamics. The significant growth and penetration of various devices come along with a tremendous increase in the number of applications supporting various domains and services which will impose a never-before-seen burden on wireless networks. The challenge however, is that each scenario requires a different wireless protocol functionality and parameter setting to optimally determine how to tune these functionalities and parameters to adapt to varying network scenarios. The traditional trial-error approach of manual tuning of parameters is not just becoming difficult to repeat but also sub-optimal for different networking scenarios. In this paper, we describe how we can leverage a deep reinforcement learning framework to be trained to learn the relation between different parameters in the physical and MAC layer and show that how our learning-based approach could help us in getting insights about protocol design optimization task.
\end{abstract}


\maketitle

\section{Introduction}

With the advent of microprocessors, the number of connected wireless devices continues to grow at a steady pace. A very recent forecast from International Data Corporation (IDC) estimates that there will be 41.6 billion connected wireless devices in 2025~\cite{report}. This significant growth and penetration of various devices come along with a tremendous increase in the number of applications supporting various domains and services. Hence, it is widely accepted that this impressive scale of devices and applications will impose a never-before-seen burden on wireless networks. To cope with the emergence of various device characteristics and application requirements, complex and custom design of high performance networking protocols is needed. Networking protocols such as WiFi and Bluetooth, traditionally, are manually designed as "general-purpose" protocols for different network characteristics and scenarios through long-time and hard-work human efforts. However, while
this approach is increasingly becoming difficult to repeat, these designed protocols are deeply rooted in inflexible, cradle-to-grave designs, and thus unable to address the demands of different network characteristics and
scenarios.
Therefore, it has now become crucial to re-engineer protocols designing process and shift toward a vision of an intelligent designing process that adapts and optimizes network protocols under various environment contexts such as device characteristics, application requirements, user objectives, and network conditions. In case of only physical layer, no single physical-layer design can work well under all scenarios, hence the natural response of the standards bodies has been to specify designs with a large number of control parameters ranging from modulation order and coding rate, to OFDM sub-carrier spacing and cyclic prefix length, to transmit power, etc.,  such that a medium can be tuned to the specific deployment scenario in the field. Each of these parameters has numerous settings leading to a large number of choices, and it becomes extremely difficult for domain experts to design a control algorithm that chooses the right algorithm depending on the scenario and the varying network conditions. 

\begin{table*}[]

\centering
\caption{Example approaches of using DL for communication protocol parameter tuning in different network stack layers}
\label{tab:examples}
\resizebox{.95\linewidth}{!}{%
\begin{tabular}{|c|c|c|c|c|c|c|} 
\hline
\textbf{\begin{tabular}[c]{@{}c@{}}Network\\ Layer\end{tabular}} & \textbf{\begin{tabular}[c]{@{}c@{}}Function/ \\ Sub-Layer\end{tabular}} & \textbf{Objective}                                                                                                                               & \textbf{\begin{tabular}[c]{@{}c@{}}Learning\\ Algorithm\end{tabular}} & \textbf{Model Input}                                                     & \textbf{\begin{tabular}[c]{@{}c@{}}Control \\ Parameter\end{tabular}}                                     & \textbf{Ref.}      \\ \hline \hline
Data Link                                                        & MAC                                                                     & \begin{tabular}[c]{@{}c@{}}Maximizing the sum throughput\\ Allocation fairness\end{tabular}                                                    & ResNet                                                                & State of Channel                                                         & Transmition of a packet                                                                                   & \cite{dlma:2019}    \\ \hline
Network                                                          & Routing                                                                 & \begin{tabular}[c]{@{}c@{}}Maximizing the minimum allocated\\  bandwidth between possible source\\ destination pairs in the network\end{tabular} & \begin{tabular}[c]{@{}c@{}}Graph-Based \\ Deep Learning\end{tabular}  & Graph of network topology                                                & \begin{tabular}[c]{@{}c@{}}Each router locally controls\\  which output interface to be used\end{tabular} & \cite{routing:2018} \\ \hline
Transport                                                        & \begin{tabular}[c]{@{}c@{}}Congestion\\ Control\end{tabular}            & \begin{tabular}[c]{@{}c@{}} Maximizing the overall utility\\ (e.g., goodput,\\ delay, $\alpha$-fairness)\end{tabular}                            & \begin{tabular}[c]{@{}c@{}}DRL\\ using\\ LSTM\end{tabular}            & \begin{tabular}[c]{@{}c@{}}States of\\ all active TCP flows\end{tabular} & Congestion window                                                                                         & \cite{DRLCC:2019}  \\ \hline
\end{tabular}}
\end{table*}
Deep learning (DL) techniques have recently been applied to various protocol and radio optimization tasks including routing~\cite{routing:2018}, congestion control~\cite{DRLCC:2019} and MAC protocol~\cite{dlma:2019}, just to name a few. Applying DL techniques can reduce manual human-based efforts to tune protocol parameters. Joseph et al.~\cite{joseph:2019} show how to design a DL-based control algorithm to jointly control two parameters namely modulation order and transmit power scaling. In their work, they show that applying DL technique may work well to control the two aforementioned parameters, but depending on the context (different devices, throughput targets, etc.,) it becomes extremely complicated to get enough insights about how black-box DL technique works, although they only tune two parameters from a large set of available control parameters. Such observations reveal why it is extremely hard for domain experts to manually design control algorithms that could capture optimal solution for each scenario.

To the best of our knowledge, the current efforts in applying DL to enhance protocol performance focus only on \textit{tuning} or \textit{controlling} protocol parameters. Table \ref{tab:examples} points to a few of the recent DL-based approaches proposed in different layers of the network stack. However, we believe that optimizing a protocol performance goes beyond individual protocol parameter \textit{tuning}. In this paper, we propose a novel Deep Reinforcement Learning (DRL)-based framework, that is not only capable of tuning protocol parameters, but also optimizing the main functionalities for each protocol. In the proposed framework, a protocol is decoupled into a set of parametric modules as DRL inputs, each representing a main protocol functionality referred as \textit{Building Blocks (BBs)}. This modularization technique helps to better understand the generated protocols and optimize the protocol design and analyze them in a systematic fashion. We feed into DRL agent, a high-level specification for a scenario, including the communication objective, the protocol BBs, measurements, and network configuration. The DRL agent then is able to learn what protocol blocks (components) are important to be included or to be neglected in the protocol design.  Therefore, this framework could provide a tool for protocol designers to re-think the blocks used in a designed protocol. In addition, our framework can be utilized as a multi-variant optimization tool that helps in alleviating the current protocol design process. When designing a protocol, domain experts should keep different application requirements, user objectives, device constraint and network conditions in mind. Considering these parameters all together is a daunting task as discussed in~\cite{joseph:2019}.

As a case study, we narrow down our focus to propose a DRL-based framework for designing MAC protocols \textit{hereafter} \deepmac. In \deepmac\ framework,  MAC protocols are decoupled into a set of parametric modules, each representing a main functionality across popular flavors of 802.11 WLANs (IEEE 802.11 a/b/g/n/ac amendments). As we showcase in Section \ref{important-blocks}, the DRL agent learns that when the load of the network is very low, it could eliminate control and sensing mechanisms (ACK and Carrier Sensing blocks, respectively) to increase the throughput of the channel by reducing the bandwidth overhead and waiting time introduced in these mechanisms. Therefore, this framework could serve as a tool for protocol designers to re-think the blocks used in a designed protocol. In addition, our framework could be utilized as a multi-variant optimization tool that helps in alleviating the current protocol design process. When designing a protocol, domain experts should keep different application requirements, user objectives, device constraint and network conditions in mind. Considering these parameters all together is a daunting task. By using this framework, domain experts provide the required specifications (objective) for a specific scenario as DRL input and could identify/capture the role that each protocol component (block) plays in varying scenarios for different objectives. It could also help domain experts to get insights about the relation between different protocol components for different objectives, although such components may not have a direct dependency/relation on each other if considered alone. 

\begin{figure*}[t!]
\centering
\begin{minipage}[t]{.5\linewidth}
	\centering
	\includegraphics[width=.99\textwidth,height=1.7 in]{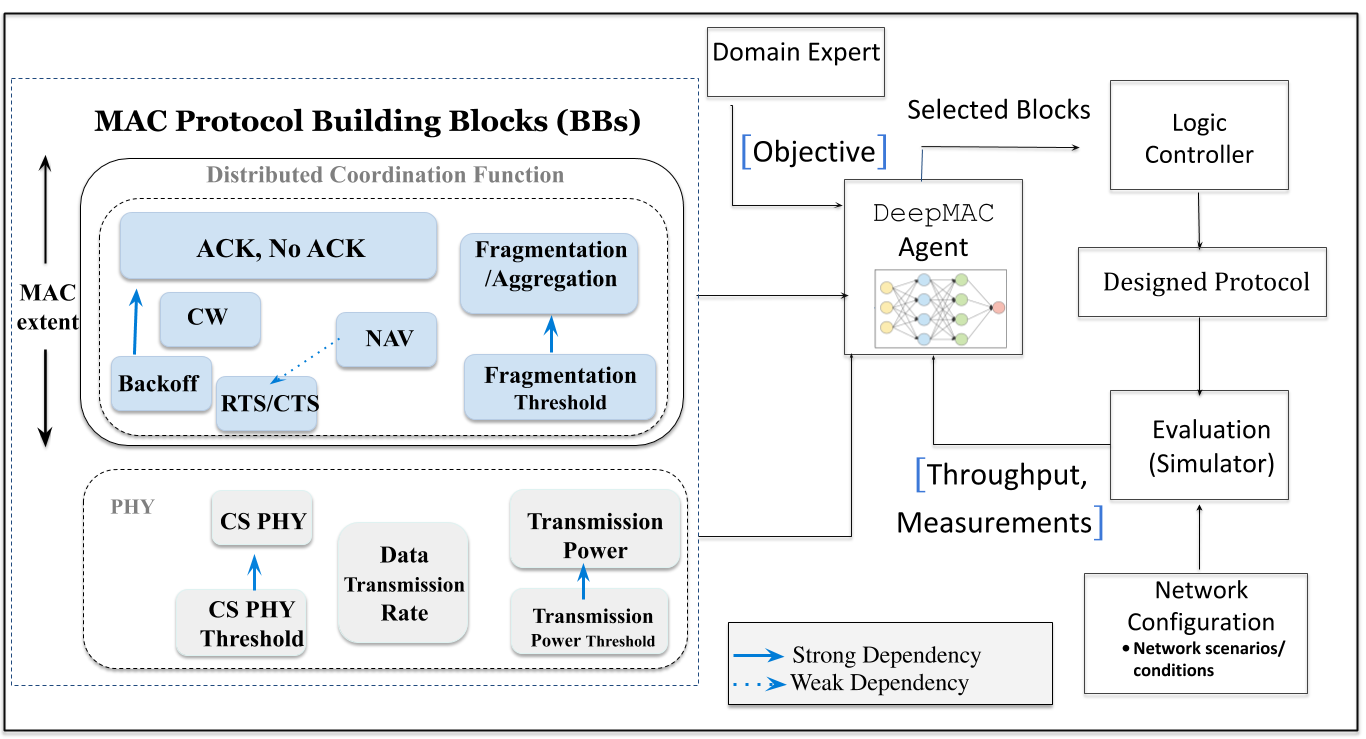}
		\vspace*{-0.1 in}
	\caption{\deepmac\ framework}
	\label{fig:framework}
	\vspace*{-0.25 in}
\end{minipage}\hfill %
\begin{minipage}[t]{0.5\linewidth}
	\includegraphics[width=.99\textwidth,height=1.7 in]{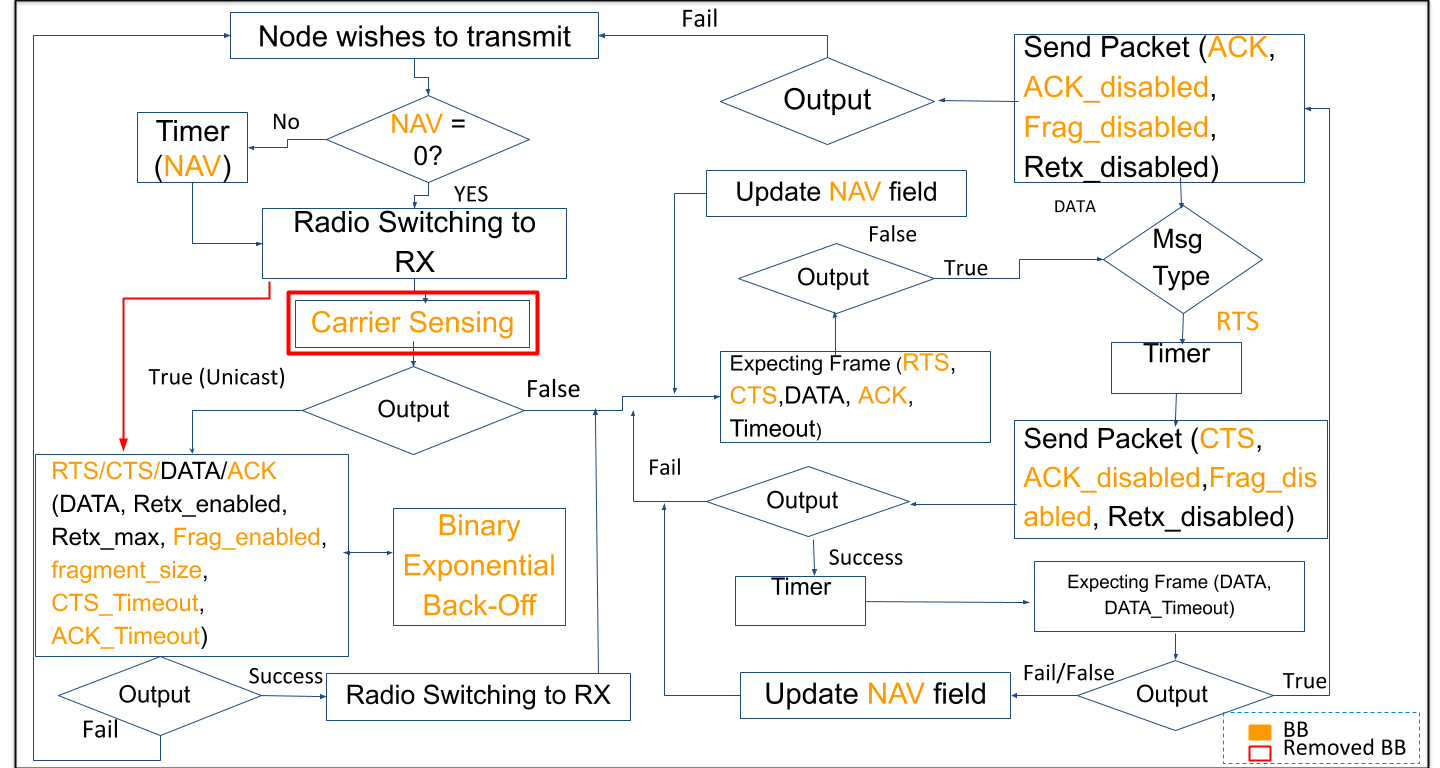}
	\centering
	\caption{ Realizing IEEE 802.11 DCF using coarse-grained MAC blocks and their corresponding parameters}
	\label{fig:DCF}
	\vspace*{-0.25 in}
\end{minipage}
\end{figure*}

\section{DeepMAC Framework}
\label{framework}
MAC protocols are required to be designed with a rich set of requirements to satisfy the needs of the overlaying applications (e.g., Augmented/ Virtual Reality, video conferencing) and scenarios. Due to the limited channel resources and a large number of devices accessing the channel, it is desirable that the MAC protocol minimizes the time wasted due to collisions or exchange of control messages.

In our previous work~\cite{pasandi:2019}, we proposed a Reinforcement Learning (RL)-based framework to optimize the MAC protocol using a simple set of functionalities. However, we discovered that the RL-based approach may face instability since the agent has to find a balance between exploration and exploitation. In~\cite{pasandi-b:2019, pasandi2019poster} we provided a complete overview of the whole protocol design framework using machine learning techniques. We described the key design considerations for the learning agent (e.g., centralized, distributed or hybrid agents) and explained how these agents should communicate with one another. We then expanded our framework~\cite{DeepMAC:2020} to use deep architecture along with new set of building blocks. In this work, we use the same framework as~\cite{DeepMAC:2020} but with the objective to get more insights about protocol design through a deeper analysis of the DRL agent. Figure \ref{fig:framework} illustrates \deepmac\ framework for optimizing the design of wireless MAC protocols. We describe the key modules of this framework in the following.

\vspace{-2mm}
\subsection{Building Blocks and Logic Controller}

\textbf{Building blocks} A network protocol is structured into several layers. Each layer is broken into a set of blocks with its own specific functionality. As described, building blocks are a set of separated parametric modular components, each of which is in charge of one (or several) specific well-defined functionality~\cite{bai:2004, doerr:2005}. The combination of different building blocks and the interactions between them determine the overall behavior of a network protocol for a given environment. In our framework, we have extracted a set of MAC protocol blocks from Wireless LAN Medium Access Control (MAC) and Physical Layer (PHY) Specifications~\cite{ieee:2016} which includes MAC functionalities across all 802.11 amendments. Figure \ref{fig:framework} shows the extracted blocks and instances of their dependencies captured based on non-Directional Multi-Gigabit (non-DMG) MAC architecture.

Once the blocks and their interactions are established, the network protocol could be represented as a graph, where the parameterized blocks are the vertices and the edges connecting the blocks represent the transition between them. Conducting the operations of individual blocks in an appropriate order, we are able to implement the protocol mechanisms. As an example, Figure~\ref{fig:DCF} shows how IEEE 802.11 DCF is realized using the extracted coarse-grained blocks. The network protocol is operated under a variety of conditions and environments, which trigger events causing the protocol to act. Therefore, when describing a building block, we should also capture the dynamic behavior of a protocol caused by different events. Building blocks should react to incoming events, conducting their main operation while interacting with each other. The dynamic behavior of a BB could be estimated if the input events are known since the behavior of the BBs is deterministic. To be exact, we could describe a building block and its dynamic behavior as the following tuple:

\begin{equation}
\label{eq:tuple}
Block: <E, P, S, F, D>
\end{equation}
Where $E$ is the \textbf{E}vent that triggers the block, $P$ is a \textbf{P}arameter inside the block that could be adjusted, $S$ is the internal \textbf{S}tate of the block, $F$ is the main \textbf{F}unction that is executed in the block, and $D$ represents the possible internal \textbf{D}ependencies between a block with other blocks. In our framework, the logic controller is responsible to check the sanity of a generated protocol. To show an example using tuple~\ref{eq:tuple}, lets consider
Backoff mechanism as a single building block. A tuple that describes this block could be: <ACK$\_$timeout, CW, Freeze/Countdown, Avoid Collision, ACK>.

\textbf{Logic Controller} In each iteration, \deepmac\ agent takes a numerical vector of building blocks, objective and network measurement as input as described in more detail later, and outputs a set of selected blocks that logic controller uses to generate the protocol. In Figure \ref{fig:DCF} example, if \deepmac\ agent decides that for the underlying scenario carrier sensing mechanism (framed by red color) should be excluded from the design, then it is the logic controller's responsibility to rewire a new version of DCF without carrier sensing. In addition, some functional blocks are dependent on each other. The logic controller is also in charge of checking the block execution sequences, their interdependencies, and interaction rules between blocks to ensure logically correct protocol design. We extracted the interdependencies between blocks from PHY and MAC specification~\cite{ieee:2016} and incorporated them into the logic controller using if-then-else rules. In our design, all dependencies are uni-directional meaning if \textit{Block A} depends on \textit{Block B} it only shows restrictions of \textit{A} $\to$  \textit{B} but not \textit{B} $\to$ \textit{A}. In the Backoff example, this mechanism is \textit{strongly} dependent on ACK block; if there is no ACK, there will be no ACK timeout to signal frame retransmission. However, ACK mechanism can be used without Backoff. Sometimes blocks are \textit{weakly} dependent on each other (see Figure \ref{fig:framework}).

\subsection{DeepMAC As A Reinforcement Learning Problem}
\deepmac\ uses RL to learn the best set of protocol blocks for different scenarios. In \deepmac, we consider a \textit{centralized} agent for the design of 802.11 MAC protocols. This centralized agent, in practice, can be based on a single supernode (e.g., the Access Point) that periodically updates its model. Meaning it decides the selected set of MAC layer blocks and parameters to be used with all other nodes in the network. The reward function can be any objective function that is required to be optimized. The \textit{reward} function in \deepmac, is the \textit{average throughput} of the link. The \textit{state} of the agent is a vector of numerical representation of the set of the building blocks, and a history with a fixed length of the average link throughput values which are used as the input to \deepmac\ agent. In this set, a value except 0 indicates that the corresponding block is included in the protocol design (each of the elements in the input vector can have different values which indicate what parameter or algorithm/method/mechanism should be used in the design), while 0 means the component is completely excluded from the design. The \textit{action} in this framework, is the act of choosing the next state among all the available states from the current state such that the reward is maximized.  

\textbf{DRL agent architecture} The neural network we have adopted is equipped with three hidden layers and an output layer. We find through experiments that this simple architecture can provide satisfactory performance, and increasing the complexity of the neural network does not contribute to performance improvements. The data is flattened before going through the hidden layers which utilize Relu as the activation function. The output layer consists of multiple neurons, each producing the Q-value of the corresponding
action.

\section{Framework Design Challenges}

In this section, we discuss essential challenges associated with designing our framework. The challenges corresponding to the main modules of the framework are discussed in the following subsections.

\subsection{Building Block Design}

Following the modular design principle in context of protocol design, two main branches exist on how to divide protocols and define the components: FSM and Data Flow. FSMs are graphical formalism that have become widely used in specifications of embedded and reactive systems. Their main drawback is that even for a moderate complicated
system, they result in large diagrams.  In order to support the dynamic design of the MAC protocols and the flexibility in selecting the optimum set of building blocks (components) by the RL agent in designing efficient protocols, capturing the interactions and dependencies between components is of crucial importance. Therefore, a desirable solution would be a flexible modular design that captures all possible interactions, dependencies, and relations of all possible design options. Yet among the main challenges to develop flexible and reusable set of building blocks is how to decide on the level of granularity of each block, and to evaluate different block granularity levels of the same function. One approach could be a brute-force in which the interactions, relations, dependencies, and conflicts between every couple of build blocks is defined by design experts. However, this approach is complex and time-consuming. Therefore, to address these issues, there should be further research on exploring how to reduce this complexity through approximation techniques, as well as whether this process could be automated. 

\subsection{RL agent design}
RL agent can be implemented in centralized or distributed approaches. Centralized agent means there is a single  agent responsible for managing the protocol design task, and then the designed protocol is enforced to be used by all the nodes in the network. Decentralized approach assumes that multiple agents perform the task of learning based on their own knowledge, including what actions to take based on the current state and expectation of other agents' actions. 
Although in a distributed approach (i.e., multi-agent environment) each agent has the flexibility to design its own optimum protocol based on its characteristics and application requirements, instability throughout the network could happen as some agents may take random actions that can affect the learning process of other agents. On the other hand, while a centralized approach is simple and easy to control and manage, it becomes computationally expensive when  number of nodes in  network grows or the state space becomes large. Moreover, a centralized approach is not suitable for heterogeneous environment where different nodes have different objectives.

Another design challenge especially for distributed approach is the communication strategy between agents in order to improve collectively their performance. The communication mechanisms between multiple agents could be mainly categorized to two main approaches; individual peer-to-peer channels and all-to-all channels. Peer-to-peer channels will enable peer agents experience similar conditions and targeting similar rewards to exchange their information with each other in order to speed-up the learning process and be able to converge to the optimum protocol within time constraints. Additionally, if the design includes a Long Short Term Memory (LSTM) that store previous experiences and observations of agents over time, an agent that is currently learning a new protocol may utilize information from another agent who dealt with similar task in the past.
\subsection{Reward Function Design} 
Another important design decision is how to design and optimize the reward function. In a global optimization, both centralized and distributed agents work towards optimizing the same goal, while in local optimization, distributed nodes can optimize their own goals. Each of these approaches have their own challenges. Different applications have different performance requirements. Therefore, defining the "right" global optimization objective is not straightforward. Optimizing the objective function relies on the assumptions that all end-hosts employ the same prescribed protocol. Thus, there is a limited support for network heterogeneity, as well as, fulfilling different applications' objectives. On the other hand, each node in a distributed optimization tries to optimize its own objective function in which it might not converge.

\section{DeepMAC Evaluation}
\label{evaluation}

\begin{table*}[t]

\begin{minipage}[t]{0.3\linewidth}\centering
\footnotesize
\centering
\caption{Simulation Configuration}
\begin{tabular}{c|c}
\textbf{Parameters}        & \textbf{Values}              \\ \hline
Frame Size                 & 1500Bytes (Default) \\
Time Slot                  & 0.2 msec                     \\
Channel Capacity           & 10 Mbps                      \\
Learning Rate ($\alpha$)   & 1                            \\
History Length ($H_t$)     & 15                           \\
Discount Factor ($\gamma$) & 0.8                         
\end{tabular}
\label{table_sim}
\end{minipage}\hfill %
\begin{minipage}[t]{.7\linewidth}\centering
\centering
\footnotesize
\caption{Blocks and their associated algorithm/ mechanism/ parameter \label{blocks}}
\begin{tabular}{|c|c|c|}
\hline
\textbf{Building Block} & \textbf{Algorithm /  Parameter}                                                        & \textbf{Default}           \\ \hline
Backoff                 & BEB, EIED                                                                              & BEB                        \\
ACK                     & \begin{tabular}[c]{@{}c@{}}No ACK,  ACK\end{tabular} & ACK                        \\
Fragmentation (Fr)           & Packet Size =200, 500, 1000 bytes                                                           & Packet Size = 1500 bytes   \\
Aggregation (Ag)             & Packet Size =2000 bytes                                                                & Packet Size =1500 bytes    \\
RTS/CTS                 & Enabled/Disabled                                                                       & N/A                    \\
CW       & 0-1023                                                                                 & $CW_{min} = 15$ \\
Carrier Sense (CS)      & Enabled/Disabled                                                                       & N/A                    \\
Data Transmission Rate (DR)  & 6/9/12/24/36/48/54 (Mbps)                                                              & 54 Mbps                    \\ \hline
\end{tabular}
\end{minipage}
\end{table*}

\begin{table*}[t]
\begin{minipage}[t]{0.3\linewidth}\centering
\footnotesize
\caption{Simulation scenarios}
\label{table_scenario}
\begin{tabular}{|c|c|c|c|}
\hline
Scenario  & Nodes & Load & Noise \\ 
\hline
1                 & 5    & Low  & No    \\
\hline
2                & 5   & Low  & Yes   \\
\hline
3              & 15  & Average  & No    \\
\hline
4              & 15   & Average  & Yes   \\
\hline
5                 & 20  & High & No    \\
\hline
6               & 20  & High & Yes   \\
\hline
7              & 50  & Saturated & No    \\
\hline
8              & 50  & Saturated & Yes   \\ 
\hline
\end{tabular}
\end{minipage}\hfill %
\begin{minipage}[t]{0.6\linewidth}\centering
\caption{Blocks selected by DeepMAC agent}
\vspace{-0.1 in}
	\includegraphics[width=.9\textwidth,height=1.62 in]{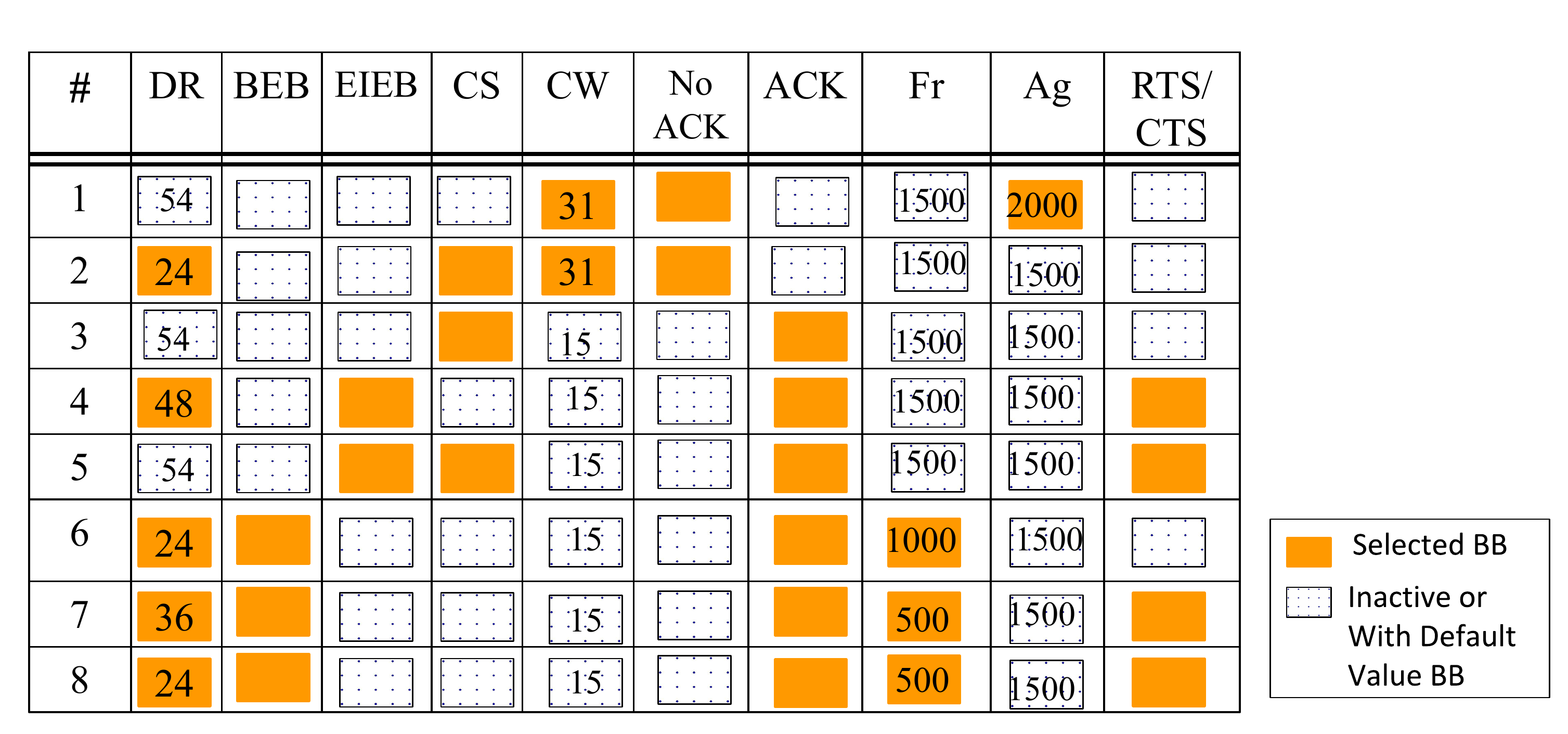}
	\label{tab:selected_components}
\end{minipage}\hfill
\end{table*}

\textbf{Performance Metrics} This section presents the numerical results and evaluation of \deepmac\ regarding \textit{block selection} by the agent under different scenarios. Before we delve into the experimental evaluation of our analysis, we clarify that we run the pre-trained DRL agent for every scenario. After training our DRL agent on a MacBook Pro with 2.9 GHz Intel Core i5 with 16 GB of memory, the agent took on average 1 ms to execute~\cite{DeepMAC:2020}. We assume that the supernode (centralized agent) uses hardware accelerators which can reduce the execution time by an order of magnitude and comfortably meet the time constraint requirements. We have not considered the convergence time of the DRL agent as a performance metric to evaluate since we have already shown the convergence of the agent in~\cite{pasandi:2019}. 

\subsection{Simulation Configuration}

We consider an ad-hoc network where individual nodes communicate with each other directly. To carry out our simulations, we use our own C++ event-driven simulator.  Table~\ref{blocks} includes the blocks and their corresponding algorithm, mechanism, or parameters that are used by \deepmac\ framework for the experiments.  Without loss of generality, we assume that each node has always a packet to transmit, and the packet generation rate follows a Poisson process. In our experiments, we consider eight different networking scenarios described in Table \ref{table_scenario}.
\subsection{Selected Blocks in Different Scenarios}
\label{important-blocks}
The selected blocks by the agent are shown in Table~\ref{tab:selected_components}. In the following, we divide our observations about DeepMAC behavior in two parts and discuss each in more detail.

\textbf{Low load with/without noise}
In scenarios with the low load when the noise is absent ( Scenario $\#1$) no control packet such as ACK or RTS/CTS is selected by the agent.  This is justifiable. Even though the control packets are much smaller than the data packets, the time spent for control packet transmission is
not negligible. 

Therefore, when the network is under-saturated, and the number of competing stations are small, the DRL agent avoids control packet overheads to maximize the throughput. Intuitively, to reduce the relative percentage of the time loss due to packet overhead and MAC coordination, frame aggregation is also selected by the agent. While for the same scenario, when the noise is present, the agent adds Career Sensing (CS) block. This could be because the agent learns such a mechanism can be useful when the throughput drops. For scenarios with the average level of noise (Scenario $\#$3,4) except common ACK mechanism selection, there is no obvious pattern. 

\textbf{High and saturated load with/without noise} We discuss the following observations for this set of scenarios: 1) The first observation for Scenario $\#$5 to 8 is the ACK mechanism selection by the agent. Intuitively, this could be because the agent learns such a mechanism can contribute to prevent more number of collisions and corresponding retransmissions to enhance the throughput. 2) When comparing scenario 5 with 6, we observe that the agent uses the Fragmentation block. The size of the sub-frames in practice plays an important factor that can influence network throughput performance for a given channel condition. The larger fragments, possibly the higher Packet Error Rate (PER) which would cause throughput drop due to a large number of retransmissions. 3) When the network is saturated, the agent selects protection mechanisms such as ACK and RTS/CTS along with smaller frame sizes and lower bitrate. However, it is not clearly obvious if the smaller frames contribute much to enhance the throughput. This is due to the fact that small fragments with the extra introduced overhead could also decrease the throughput performance. The varying results reveal why it is extremely hard for an algorithm based on manually-specified rules and thresholds to capture the optimal solution, and
why it is instead better to use such a deep learning-based tool to
optimize the design of control algorithms and get insights about what functionality is useful under what scenario.

\section{Conclusion}
 
In this paper, we proposed and evaluated a framework for MAC protocol design optimization using a DRL-based approach. We have shown that by observing the decisions of the \deepmac\ agent and using a method such as input modularization (protocol decomposition into building blocks), it is possible to extract information about the associated component selection by the agent. We envision this method could offer useful insights, especially to protocol designers to build deeper perception about the significance of an individual or a set of protocol blocks (functions) under different scenarios. This could help them focusing on enhancements/ modifications of important components than focusing on the whole protocol performance in order to enhance the protocol design and performance. 

\bibliographystyle{ACM-Reference-Format}
\bibliography{references}

\end{document}